\documentstyle[twoside,fleqn,espcrc2]{article}

\newcommand{\AmS}{{\protect\the\textfont2
  A\kern-.1667em\lower.5ex\hbox{M}\kern-.125emS}}

\def\beq{\begin{equation}}
\def\eeq{\end{equation}}
\def\bea{\begin{eqnarray}}
\def\eea{\end{eqnarray}}
\def\bq{\begin{quote}}
\def\eq{\end{quote}}

\def\nnb{\nonumber}
\def\ga{\left(}
\def\dr{\right)}

\def\lb{\lbrack}
\def\rb{\rbrack}
\def\rar{\rightarrow}
\def\lrar{\Longrightarrow}
\def\nnb{\nonumber}
\def\la{\langle}
\def\ra{\rangle}
\def\nin{\noindent}
\def\ba{\begin{array}}
\def\ea{\end{array}}

\def\b{\bullet}

\title{Masses, Decays and Mixings of Gluonia in QCD }
\author{ Stephan Narison\address{
Laboratoire de Physique Math\'ematique,
Universit\'e de Montpellier 2\\
Place Eug\`ene Bataillon,
34095 - Montpellier Cedex 05, France.\\
E-mail: narison@lpm.univ-montp2.fr }}
\begin{document}
\pagestyle{plain}
\begin{abstract}
\noindent
This is a review of the estimate of the gluonia
masses, decay and mixings from QCD
spectral sum rules and low-energy theorems. Some phenomenological 
maximal gluonium-quarkonium
 mixing shemes in the
scalar sector are presented. This talk is a compact version of
the work in Ref. \cite{SN}.
\end{abstract}
\maketitle
\section{Introduction}
\nin
In addition to the well-known
mesons and baryons, one of the main consequences of the non-perturbative aspects
of the QCD theory is the possible existence of the gluon bound states 
(gluonia or glueballs) or/and of a gluon continuum.
Since the pionneering work of Fritzsch and Gell-Mann \cite{FRITZ}, 
a lot of theoretical and experimental efforts have been devoted to 
study the gluonia properties. 
In this talk, we present
an update of the predictions for the masses, decay constants and
mass-mixing from QCD spectral sum rules (QSSR) \`{a} la
SVZ \cite{SVZ} \footnote{For a recent review on the sum rules, 
see e.g. \cite{SNB}}. We also report some low-energy theorem (LET) 
\cite{VENEZIA} predictions for
the widths of unmixed scalar gluonia and present (almost)
complete mixing schemes for explaining the complicated spectra
of the observed scalar resonances below 2 GeV. 
\section {The gluonic currents}
\nin
In this paper, we shall consider the lowest-dimension gauge-invariant
gluonic currents
that can be built from two gluon fields:
\bea
J_s&=& \beta(\alpha_s) G_{\alpha\beta}G^{\alpha\beta},\nnb\\
\theta^g_{\mu\nu}&=&-G_{\mu}^{\alpha}G_{\nu\alpha}+\frac{1}{4}g_{\mu\nu}
G_{\alpha\beta}G^{\alpha\beta}, \nnb\\
Q(x)&=&\ga \frac{\alpha_s}{8\pi}\dr \mbox{tr}~G_{\alpha\beta}
\tilde{G}^{\alpha\beta},
\eea
and three-gluon ones:
\beq
J_{3G}=g^3f_{abc}G^aG^bG^c
\eeq
where the sum over colour is understood; 
$\beta(\alpha_s)$ is the QCD $\beta$-function, while
$\tilde{G}_{\mu\nu}\equiv(1/2)\epsilon_{\mu\nu\alpha\beta}{G}^{\alpha\beta}$,
which have respectively the 
quantum numbers of the $J^{PC}= 0^{++},~2^{++}$ and $0^{-+}$ 
gluonia for the two-gluon fields, 
and to the $0^{++}$ one for the three-gluon fields. The former two enter 
into the QCD energy-momentum tensor
$\theta_{\mu\nu}$, while the third one is the U(1)$_A$ axial-anomaly current. 
\section{QCD spectral sum rules}
\nin
The analysis of the gluonia masses and couplings will be
done using the method of
QSSR. In so doing, we shall work with the generic 
two-point correlator: 
\beq
\psi_G(q^2) \equiv i \int d^4x ~e^{iqx} \
\la 0\vert {\cal T}
J_G(x)
\ga J_G(0)\dr ^\dagger \vert 0 \ra ,
\eeq
built from the previous gluonic currents $J_G(x)$, which obeys the well-known
K\"allen--Lehmann dispersion relation:
\beq
\psi_G (q^2) = 
\int_{0}^{\infty} \frac{dt}{t-q^2-i\epsilon}
~\frac{1}{\pi}~\mbox{Im}  \psi_G(t) ~ + ...,
\eeq
where ... represent subtraction points, which are
polynomials in the $q^2$-variable. This $sum~rule$
expresses in a clear way the {\it duality} between the integral involving the 
spectral function Im$ \psi_G(t)$ (which can be measured experimentally), 
and the full correlator $\psi_G(q^2)$, 
which can be calculated directly in
QCD.
\subsection{The two-point correlator in QCD}
\nin
In addition to the usual perturbative contribution from the $bare$ loop,
the non-perturbative contributions can be parametrized by the vacuum
condensates of higher and higher dimensions in the Wilson expansion \cite{SVZ}
\footnote{In the present analysis, we shall limit ourselves
to the computation of the gluonia masses
in the massless quark limit $m_i=0$.}:
\bea
\psi_G(q^2)
\simeq &&\sum_{D=0,2,4,...}\frac{1}{\ga -q^2 \dr^{D/2}}\nnb\\ 
&\times&\sum_{dim O=D} C^{(J)}(q^2,\nu)\la O(\nu)\ra,
\eea
provided 
that $-q^2$ is much greater than $\Lambda^2$;
$\nu$ is an arbitrary scale that separates the long- and
short-distance dynamics; $C^{(J)}$ are the Wilson coefficients calculable
in perturbative QCD by means of Feynman diagrams techniques.
\\
$\b$ The dominant condensate contribution 
in the chiral limit $m_i=0$ is due to
the dimension-four gluonic condensate $\la\alpha_s G^2 \ra$,
introduced by SVZ \cite{SVZ}, and which has been estimated recently from the
$e^+e^-\rar~I=1$ hadron data \cite{SNL} and from the heavy
quark-mass splittings \cite{SNH}:
\beq
 \la\alpha_s G^2 \ra \simeq (0.07\pm 0.01)~\mbox{GeV}^4.
\eeq
\nin
$\b$ The first non-leading contribution comes from the triple gluon condensate
$gf_{abc}\la G^aG^bG^c \ra$, whose direct extraction from
the data is still lacking. We use its approximate 
value from the dilute  gas instanton model \cite{SVZ}:
 \beq
 g^3f_{abc}\la G^aG^bG^c \ra\approx (1.5\pm 0.5) \mbox{GeV}^2)\la\alpha_s G^2 \ra~,
\eeq
within a factor 2 accuracy.\\
\nin
$\b$ In addition to these terms, the UV renormalon and some eventual other
effects induced by the resummation of the QCD series, and not included in the OPE, 
can contribute to the correlator
as a term of dimension 2 \cite{ZAK}.We consider that their effects can be safely 
taken into account in the estimate of the
errors from the last known term of the truncated perturbative series 
\footnote{See e.g. the
estimate of the errors in the determination of
$\alpha_s$ from $\tau$ decays \cite{SETTLES}.}. \\
\nin
$\b$ It has also been argued 
\cite{NSVZ}, using the
dilute gas approximation, that in the gluonia channels, instanton plus 
anti-instanton effects manifest themselves
as higher dimension $(D=11)$ operators.
However, at the scale (gluonia scale) where the following sum rules
are optimized, which is much higher
than the usual case of the $\rho$ meson, 
we can safely omit these terms \footnote {Their 
quantitative estimate is quite inaccurate because of
 the great sensitivity of the result on
the QCD scale $\Lambda$, and on some other less controllable parameters and
coefficients.}, like any other 
higher-dimensional operators beyond $D=8$.\\
Throught this paper, we shall use for three active flavours,
the value of the QCD scale \cite{BETHKE}:
\beq
\Lambda= (375\pm 125)~\mbox{MeV}.
\eeq 
\subsection{The spectral function and its experimental measurement}
\nin
It can be best illustrated
in the case of the flavour-diagonal 
light quark vector current, where the spectral function Im$\Pi$(t) can
be related to the $e^+e^-$ into $I=1$ hadrons data via the optical
theorem as:
\beq
\sigma (e^+e^-\rar\mbox{hadrons})=
\frac{4\pi^2\alpha}{t}e^2~\frac{1}{\pi}~
\mbox{Im}  \Pi(t), 
\eeq
or, within a vector meson dominance assumption, to the leptonic width of 
the $\rho$ resonance:
\beq
\Gamma_{\rho\rar e^+e^-}\simeq \frac{2}{3}\pi \alpha^2\frac{M_\rho}
{2\gamma^2_\rho},
\eeq
via the meson coupling to the electromagnetic current:
\beq
\la 0|J^\mu|\rho\ra =\frac{M^2_\rho}{2\gamma_\rho}\epsilon^\mu.
\eeq
More generally, the resonance contribution to the spectral function can be
introduced,
via its decay constant $f_G$ analogous to $f_\pi=93.3$ MeV:
\beq
\la 0|J_G|G\ra =\sqrt{2}f_GM_G^2~...,
\eeq
where ...~represent the Lorentz structure of the matrix elements.
\subsection{The form of the sum rules} 
\nin
The 
previous dispersion relation can be improved from the 
uses of
an infinite number of derivatives and infinite values of $q^2$, 
but keeping their ratio fixed as $\tau \equiv n/q^2$. In this
way, one obtains the Laplace \footnote{Also called Borel or exponential.} 
sum rules \cite{SVZ,BB,SR} \footnote{QCD finite energy 
sum rules (FESR) \cite{LAUNER} lead to equivalent results and
complement the Laplace sum rules.}:
\beq
{\cal L}_G(\tau)
= \int_{t_\leq}^{\infty} {dt}~\mbox{exp}(-t\tau)
~\frac{1}{\pi}~\mbox{Im} \psi_G(t),
\eeq
where $t_\leq$ is the hadronic threshold. The advantage of this sum 
rule with respect to the previous dispersion relation is the
presence of the exponential weight factor, which enhances the 
contribution of the lowest resonance and low-energy region
accessible experimentally. For the QCD side, this procedure has
eliminated the ambiguity carried by subtraction constants
(arbitrary polynomial
in $q^2$), and has improved the convergence of
the OPE by the presence of the factorial dumping factor for each
condensates of given dimensions. 
The ratio of sum rules:
\beq
{\cal R}_G \equiv -\frac{d}{d\tau} \log {{\cal L}_G},
\eeq
 or its slight modification, is a useful quantity to work with,
 in the determination of the resonance mass, as it is equal to the 
mass squared, in  
 the simple duality ansatz parametrization
\footnote{As tested in the meson channels where complete data are available, this
parametrization gives a good description of the spectral integral for 
the sum rule analysis.}:
\bea
&&``\mbox{one resonance}"\delta(t-M^2_R)
 \ + \ \nnb\\
&& ``\mbox{QCD continuum}" \Theta (t-t_c),
\eea
of the spectral function, where the
resonance enters by its coupling to the quark current; 
$t_c$ is
the continuum threshold which is, like the 
sum rule variable $\tau$, an  a priori arbitrary 
parameter. 
\subsection{Conservative optimization criteria}
\nin
Different optimization criteria are proposed in the literature,
which, to my opinion, complete one another, if used carefully.
The {\it sum rule window} of SVZ is a compromise region
where, at the same time, the OPE makes sense while the spectral
integral is still dominated by the lowest 
resonance. This is indeed
satisfied when the Laplace sum rule presents a minimum in
$\tau$, where there is an equilibrium between the
non-perturbative and high-energy region effects.
However, this criterion is not yet sufficient as the value of this
minimum in $\tau$ can still be greatly affected by the value
of the continuum threshold $t_c$. 
The needed extra condition is to find the region where
the result has also a minimal sensitivity on the change of the
$t_c$ values ($t_c$ stability).
The $t_c$ values obtained in this way are about the same as the one from
the so-called heat evolution test of the local duality 
FESR \cite{LAUNER}. However, in some cases, this $t_c$
value looks too high, compared with the mass of the
observed radial excitation, and the procedure
tends to overestimate the predictions. More precisely, the result obtained
in this way can be considered as a phenomenological
upper limit.
Therefore, in order to have a {\it conservative} prediction from 
the sum rules method, one can consider the value of $t_c$ at which one
starts to have a $\tau$-stability up to where one
has a $t_c$ stability. In case there is no $t_c$ stability
nor FESR constraint on $t_c$, one can consider that the
prediction is still unreliable. In this paper, we shall limit ourselves to
extracting the results satisfying the $\tau$ (Laplace)
 and $t_c$ stability criteria.\footnote{Many results in the literature
on QCD spectral 
sum rules are obtained using  only the first condition.}  
\begin{table*}[hbt]
\setlength{\tabcolsep}{1.3pc}
\newlength{\digitwidth} \settowidth{\digitwidth}{\rm 0}
\catcode`?=\active \def?{\kern\digitwidth}
\caption{ Unmixed gluonia masses and couplings from QSSR.}
\begin{tabular}{c c c c c c}
\hline 
$J^{PC}$&Name&{Mass [GeV]}&&$f_G$ [GeV]&$\sqrt{t_c}$ [GeV]\\
&&Estimate&Upper Bound&&\\
\hline 
$0^{++}$&$G$&$1.5\pm 0.2$&$2.16\pm 0.22$&$390\pm 145$&$2.0\sim 2.1$\\
&$\sigma_B$&1.00 (input)&&1000&\\
&$\sigma'_B$&1.37 (input)&&600&\\
&$3G$&3.1&&62&\\
$2^{++}$&$T$&$2.0\pm 0.1$&$2.7\pm 0.4$&$80\pm 14$&$2.2$\\
$0^{-+}$&$P$&$2.05\pm 0.19$&$2.34\pm 0.42$&$8\sim 17$&$2.2$\\
&$E/\iota$&1.44 (input)&&$ 7 ~:~J/\psi\rar\gamma\iota$&\\
\hline 
\end{tabular}
\end{table*}
\section{Masses and decay constants of the unmixed gluonia}
\nin
The different expressions of the sum rules for each channels have been given
in \cite{SN}. Applying the previous stability criteria, we obtain the spectra
given in Table 1.
Our results
satisfy the mass hierarchy $M_S<M_P\approx M_T$, which suggests that the
scalar is the lightest gluonium state as also expected from lattice calculations
\cite{TEPER} and QCD inequalities \cite{WEST}. However, 
the consistency of the different subtracted and unsubtracted
sum rules in the scalar sector requires
the existence of an additional lower mass and broad
$\sigma$-meson coupled strongly both to gluons and
to pairs of Goldstone bosons (similar to the $\eta'$ of the $U(1)_A$ channel), 
whose effects can be
missed in a one-resonance parametrization of the spectral function, and
in the present lattice quenched approximation. One should also notice that 
the values of $\sqrt{t_c}$,
which are about the mass of the next radial excitations, indicate that the
mass-splitting between the ground state and the radial excitations is relatively much
smaller ($30\%$) than in the case of ordinary hadrons (about $70\%$ for the $\rho$ meson), such
that one can expect rich gluonia spectra in the vicinity of 2--2.2 GeV, in addition to the
ones of the lowest ground states. The upper 
bounds on the gluonium mass squared given in Table 1
have been obtained from the minimum (or
inflexion point) of the ratios of sum rules, after using the positivity of 
the spectral functions.
\section{Natures of the $\zeta(2.2)$ and $E/\iota$(1.44)}
\begin{table*}[hbt]
\setlength{\tabcolsep}{.65pc}
\caption{ Unmixed scalar gluonia and quarkonia decays}
\begin{tabular}{ c c c c c c c c}
\hline 
Name&Mass&$\pi^+\pi^-$&$K^+K^-$&$\eta\eta$
&$\eta\eta'$&$(4\pi)_S$&$\gamma\gamma$\\ 
& [GeV]& [GeV]&[MeV]&
[MeV]&[MeV]&[MeV]&[keV]\\ 

\hline 
{Gluonia}&&&&&&&\\
$\sigma_B$&$0.75\sim 1.0$&$ 0.2\sim 0.5$&$SU(3)$&$SU(3)$&&&$0.2\sim 0.3$\\
&(input)&&&&&\\
$\sigma'_B$&1.37&$0.5\sim 1.3$&$SU(3)$&$SU(3)$&&$43\sim 316$&$0.7\sim 1.0$\\
&(input)&&&&&(exp)&\\
$G$&1.5&$\approx 0$&$\approx 0$&$1.1\sim 2.2$&$5\sim 10$&$60\sim 138$&$0.2\sim 1.8$\\
{Quarkonia}
&&&&&&&\\
$S_2$&1.&0.12&$SU(3)$&$SU(3)$&&&0.67\\
$S'_2$&$1.3\approx \pi'$&$0.30\pm 0.15$&$SU(3)$&$SU(3)$&&&$4\pm 2$\\
$S_3$&$1.47\pm 0.04$&&$73\pm 27$&$15\pm 6$&&&$0.4\pm 0.04$\\
$S'_3$&$\approx 1.7$&&$112\pm 50$&$SU(3)$&&&$1.1\pm 0.5$\\
\hline
\end{tabular}
\end{table*}
\nin
$\b$ The $\zeta(2.2)$ is a good $2^{++}$ gluonium candidate because of its mass
(see Table 1) \footnote{The small quarkonium-gluonium (mass) mixing angle \cite{BRAMON}
allows to expect that the observed meson mass is about the same as the one
in Table 1} and small width in $\pi\pi$ ($\leq$ 100 MeV). However,
the value of $t_c$ can suggest that
the radial excitation state is also in the 2 GeV region, which
should stimulate further experimental searches.\\
$\b$ The $E/\iota$ (1.44) or other particles in this region \cite{LANDUA} is
too low for being the lowest pseudoscalar gluonium. One of these states are
likely to be the first radial excitation of the $\eta'$ as its coupling to the
gluonic current is weaker than the one of the $\eta'$ and of the
gluonium (see Table 1).
\section{Decay widths of the scalar gluonia}
\subsection{$\sigma_B$ and $\sigma'_B$ couplings to $\pi\pi$}
\nin
For this purpose, we consider the vertex:
\beq
V(q^2)=\la\pi_1|\theta^\mu_\mu|\pi_2\ra,~~~~~q=p_1-p_2~,
\eeq
where:
$V(0)=2m^2_\pi~$. 
In the chiral limit $(m^2_\pi \simeq 0)$, 
the vertex obeys the dispersion relation:
\beq
V(q^2)=\int_0^\infty \frac{dt}{t-q^2-i\epsilon}
~\frac{1}{\pi}\mbox{Im} V(t),
\eeq
which gives the 1st NV sum rule \cite{VENEZIA}:
\beq
\frac{1}{4}\sum_{S\equiv\sigma_B,\sigma'_B,G}g_{S\pi\pi}\sqrt{2}f_S \simeq 0.
\eeq
Using the fact that $ V^{\prime} (0)=1$ \cite{NSVZ2}, one obtains 
the second NV sum rule:
\beq
\frac{1}{4}\sum_{S\equiv\sigma_B,\sigma'_B,G}g_{S\pi\pi}\sqrt{2}f_S/M^2_S=1.
\eeq
Identifying the $G$ with the $G(1.5\sim 1.6)$ at GAMS (an almost pure gluonium
candidate), 
we can neglect then its coupling to $\pi\pi$, and deduce:
\bea
g_{\sigma_B\pi\pi}&\approx& \frac{4}{\sqrt{2}f_{\sigma_B}}~\frac{1}
{\ga 1- 
M^2_{\sigma_B}/M^2_{\sigma'_B}\dr}\nnb\\
g_{\sigma'_B\pi\pi}&\approx&g_{\sigma_B\pi\pi}
\ga \frac{f_{\sigma_B}}{f_{\sigma'_B}}\dr.
\eea
Using $M_{\sigma'_B}
\approx 1.37$ GeV, one can deduce the width into $\pi\pi~ (\pi^+\pi^-$ and $2\pi^0)$
given in Table 2 
\footnote{We have checked in \cite{SN} that 
finite-width corrections do not change the result, while
a very light $\sigma_B$ around $500$ MeV 
cannot be broad. We use
the normalization:
$$
\Gamma(\sigma_B\rar\pi\pi)=\frac{|g_{\sigma_B\pi\pi}|^2}
{16\pi M_{\sigma_B}}\ga{1-\frac{4m^2_\pi}{M^2_{\sigma_B}}}\dr^{1/2}.
$$ }.
 Our result indicates 
the presence of gluons inside the wave 
functions of the broad $\sigma_B$ resonance below 1 GeV and of the $\sigma'(1.37)$, which 
can decay copiously into $\pi\pi$ \footnote{The decays of the physically 
observed
states will be discussed later on.}. 
\subsection{{\it G}(1.5) coupling to $\eta$$\eta'$}

\nin
Analogous low-energy theorem (see NV) gives:
\beq
\la \eta_1|\theta^\mu_\mu|\eta_1\ra = 2M^2_{\eta_1},
\eeq
where $\eta_1$ is the unmixed $U(1)$ singlet state of mass
$M_{\eta_1}\simeq $ 0.76 GeV \cite{WITTEN}.
Writing the dispersion relation for the vertex, one obtains the NV 
sum rule:
\beq
\frac{1}{4}\sum_{S\equiv\sigma_B,\sigma'_B,G}g_{S\eta_1\eta_1}\sqrt{2}f_S=
2M^2_{\eta_1},
\eeq
which, by assuming a $G$-dominance of the
vertex sum rule, leads to:
\beq \label{coup}
g_{G\eta_1\eta_1}\approx (1.2\sim 1.7)~\mbox{GeV}.
\eeq
Introducing the ``physical" $\eta'$ and $\eta$ through:
\bea 
\eta'\sim \cos\theta_P \eta_1-\sin\theta_P \eta_8\nnb\\
\eta\sim \sin\theta_P \eta_1+\cos\theta_P \eta_8,
\eea
where \cite{PDG,GILMAN}
$\theta_P\simeq -(18\pm 2)^\circ $ is the pseudoscalar mixing angle,
one obtains the width given in Table 2.
The previous scheme is also known to predict (see NV and \cite{GERS}):
\beq
r\equiv \frac{\Gamma_{G\eta\eta}}{\Gamma_{G\eta\eta'}}\simeq 0.22,~~~~~
g_{G\eta\eta}\simeq \sin\theta_Pg_{G\eta\eta'},
\eeq
compared with the GAMS data \cite{LANDUA} $r\simeq 0.34\pm 0.13$, and
which implies the width $\Gamma_{G\eta\eta}$ in Table 2.
This result can then suggest that the $G(1.6)$ seen by the GAMS group is a 
pure gluonium, which
is not the case of the particle seen by Crystal Barrel \cite{LANDUA}
which corresponds to $r\approx 1$.
\subsection{$\sigma'_B(1.37)$ and {\it G}(1.5) couplings 
to $4\pi$}
\nin
Within our scheme, we expect that the $4\pi$ are mainly $S$-waves initiated 
from the decay of pairs of $\sigma_B$. Using:
\beq
\la \sigma_B|\theta^\mu_\mu|\sigma_B\ra = 2M^2_{\sigma_B},
\eeq
and writing the dispersion relation for the vertex, one obtains the sum 
rule:
\beq
\frac{1}{4}\sum_{i=\sigma_B,\sigma'_B,G}g_{S\sigma_B\sigma_B}\sqrt{2}f_S=
2M^2_{\sigma_B}.
\eeq
We identify the $\sigma'_B$ with the observed $f_0(1.37)$, and use
its observed width into $4\pi$, which is about $(46\sim 316)$ {MeV}
\cite{PDG,LANDUA} ($S$-wave part).
Neglecting, to a first approximation, the $\sigma_B$ contribution
to the sum rule, we can deduce:
\beq
g_{G\sigma_B\sigma_B}\approx (2.7\sim 4.3)~\mbox{GeV},
\eeq
which leads to the width of 60-138 {MeV}, 
much larger than the one into $\eta\eta$ and $\eta\eta'$
in Table 2.
This feature seems to be satisfied
by the states seen by GAMS and Crystal Barrel. Our previous approaches show 
the consistency in
interpreting the $G(1.6)$ seen at GAMS as an ``almost" pure gluonium state
(ratio of the $\eta\eta'$ versus the $\eta\eta$ widths), 
while the
state seen by the Crystal Barrel, though having a gluon component 
in its wave function,
cannot be a pure gluonium because of its prominent
decays into $\eta\eta$ and $\pi^+\pi^-$. We shall see later on
that the Crystal Barrel state
can be better explained from a mixing of the GAMS gluonium with the $S_3(\bar ss)$
and $\sigma'_B$ states.
\subsection{$\sigma_B$, $\sigma'_B$ and $G$ couplings to $\gamma\gamma$}
\nin
The two-photon widths of the $\sigma_B,~\sigma'_B$ and $G$
 can be obtained by identifying the 
Euler-Heisenberg effective Lagrangian \cite{NSVZ2}
\footnote{$F^{\mu\nu}$ is the photon field strength,
$Q_q$ is the quark charge in units of $e$, $-\beta_1=9/2$ for three
 flavours, and $m_q$ is the ``constituent" quark mass, which we shall
take to be
$
m_u\simeq m_d \simeq M_\rho/2,~~~~m_s\simeq M_\phi/2~.
$ }:
\bea
{\cal L}_{\gamma g}&=&\frac{\alpha\alpha_sQ^2_q}{180m^2_q}\lb 
28F_{\mu\nu}F_{\nu\lambda}G_{\lambda\sigma}G_{\sigma\mu}\nnb\\
&+&14F_{\mu\nu}G_{\nu\lambda}F_{\lambda\sigma}G_{\sigma\mu}
-F_{\mu\nu}G_{\mu\nu}F_{\alpha\beta}G_{\alpha\beta}\nnb\\ &-&
F_{\mu\nu}F_{\mu\nu}G_{\alpha\beta}G_{\alpha\beta}\rb,
\eea
with the scalar-$\gamma\gamma$ Lagrangian
\beq
{\cal L}_{S\gamma\gamma}=g_{S\gamma\gamma}\sigma_B(x)
F^{(1)}_{\mu\nu}F^{(2)}_{\mu\nu}.
\eeq


\nin
This leads to the sum rule:
\beq
{g_{S\gamma\gamma}}\simeq
\frac{\alpha}{60}\sqrt{2}{f_{S}M^2_{S}}
\ga\frac{\pi}{-\beta_1}\dr\sum_{q\equiv u,d,s}{\frac{Q^2_q}{m_q^4}},
\eeq
from which we deduce the couplings
\footnote{Here and in the following, we shall use
$M_{\sigma_B}\approx (0.75\sim 1.0)$ GeV.}:
\bea
g_{S\gamma\gamma}&\approx &(0.4\sim 0.8)\alpha~\mbox{GeV}^{-1},
\eea
($S\equiv \sigma_B,
~\sigma'_B,~ G$) and the widths in Table 2,
 smaller (as expected) than the  
well-known quarkonia width:
$
\Gamma(f_2\rar\gamma\gamma)\simeq 2.6
$ keV.
Alternatively,  one can use the trace anomaly:
$\la 0|\theta^\mu_\mu|\gamma_1\gamma_2\ra$ and the 
fact that its RHS is ${\cal O}(k^2)$, in order to get the sum rule \cite{ELLIS,NSVZ}
($R\equiv 3\sum Q^2_i$):
\beq
\la 0|\frac{1}{4}\beta(\alpha_s)G^2|\gamma_1\gamma_2\ra \simeq -
\la 0|\frac{\alpha R}{3\pi}F^{\mu\nu}_1F^{\mu\nu}_2|\gamma_1\gamma_2\ra,
\eeq
from which one can deduce the couplings:
\beq
\frac{\sqrt{2}}{4}\sum_{S\equiv\sigma_B,\sigma'_B,G}f_Sg_{S\gamma\gamma}\simeq 
\frac{\alpha R}{3\pi}.
\eeq
It is easy to check that the previous values of the couplings also
satisfy the trace anomaly sum rule.
\subsection{$J/\psi\rar\gamma S$ radiative decays}
\nin
As stated in \cite{NSVZ2}, one can estimate this process, using dipersion relation
techniques, by saturating the spectral function by the $J/\psi$ plus a
continuum. The glue part of the amplitude can be converted into a physical
non-perturbative
matrix element $\la 0|\alpha_s G^2|S\ra$ known through the decay constant $f_S$
estimated from QSSR.
By assuming that the continuum is small, one obtains 
\footnote{We use $M_c\simeq 1.5$ GeV for the charm constituent quark mass and 
$-\beta_1=7/2$ for six flavours.}:
\bea
\Gamma(J/\psi\rar\gamma S)\simeq&& \frac{\alpha^3\pi}{\beta_1^2 656100}
\ga\frac{M_{J/\psi}}{M_c}\dr^4\ga\frac{M_{S}}{M_c}\dr^4\nnb\\ 
&&\frac{\ga 1-M^2_S/M^2_{J/\psi}\dr^3}{\Gamma(J/\psi\rar e^+e^-)}f^2_S.
\eea
This leads to (in units of $10^{-3}$) \footnote{
From the previous results, one can also deduce the corresponding
stickiness defined in \cite{CHAN}.}:
\bea
B(J/\psi\rar\gamma S)&\times& B(S\rar~ \mbox{all})
\approx 0.4\sim 1.
\eea
for $S\equiv \sigma_B,~
\sigma'_B,~ G$.
These branching ratios can be compared with the observed 
$B(J/\psi\rar\gamma f_2)\simeq 1.6\times 10^{-3}$. The $\sigma_B$ could already have been
 produced, but might
have been confused with the $\pi\pi$
background. The ``pure gluonium"
$G$ production rate is relatively small,
contrary to the na\"{\i}ve expectation for a glueball production. In our
approach, this is due to the relatively small value of its decay constant,
which controls the non-perturbative dynamics. 
Its observation from this process should wait for the $\tau$CF machine. 
However, we do not exclude the possibility that a state resulting from a 
quarkonium-gluonium mixing may be produced at higher rates. 
\section{Properties of the scalar quarkonia}
\subsection{Mass and decay constants}
\nin
We shall consider the $SU(2)$ singlet $S_2(\bar uu+\bar dd)$
and the $SU(3)$ $S_3(\bar ss)$ states.
We consider the former state as the $SU(2)$ partner of the $a_0(0.98)$ associated 
to the divergence of the charged vector current of current algebra:
\beq
\partial_\mu V^\mu(x)\equiv (m_u-m_d)\bar u(i\gamma_5)d.
\eeq
We expect from the good realization of the $SU(2)$ symmetry that they
are degenerate in mass,
where we shall use the QSSR prediction \cite{SNB}:
\beq
M_{a_0}\simeq (1\sim 1.05)~\mbox{GeV},
\eeq
in good agreement with the observed $a_0$ mass.
The continuum threshold at which the previous prediction has been 
optimized can 
roughly indicate the mass of the next radial excitation, which is 
about the $f_0(1.37)$ mass \cite{SNB}:
\beq
M_{S'_2}\approx \sqrt{t_c}\simeq (1.1\sim 1.4) ~\mbox{GeV}\approx M_{\pi'}.
\eeq
In order to compute the mass of the $S_3(\bar ss)$ state,
we work with the double ratio of Laplace transform sum rules:
\bea
\frac{{\cal R}_{\bar ss}}{{\cal R}_{\bar us}}&\simeq& \frac{M^2_{\bar ss}}
{M^2_{K^*_0 (1.43)}}, \eea
where ${\cal R}_{\bar qs}$ has been defined in Eq. (14) and corresponds to the 
the two-point correlator:
\beq
\psi_{\bar qs}(q^2) \equiv i \int d^4x ~e^{iqx} \
\la 0\vert {\cal T}
J_{\bar qs}(x)
\ga J_{\bar qs}(0)\dr ^\dagger \vert 0 \ra ,
\eeq
associated to the scalar current:
\beq
J_{\bar qs}(x)=(m_q+m_s)\bar qs,~~~~~~~~~~~~~q\equiv d,s.
\eeq
At the stability point, one obtains \footnote{
We have used $\overline{m}_s(1$GeV)~$\simeq (150\sim 190)$ MeV correlated 
to the values of $\Lambda$ \cite{SNM}.
}:
\bea
M_{\bar ss}/M_{K^*_0 (1.43)}&\simeq& 1.03\pm 0.02
\lrar \nnb\\
M_{\bar ss} &\simeq& (1474\pm 44)~\mbox{MeV},
\eea
confirming the earlier QSSR estimate in \cite{SNB}.
The result indicates the mass hierarchy:
\beq
M_{S_2\equiv \bar uu+\bar dd}
<M_{K^*_0\equiv \bar us}<M_{\bar ss}.
\eeq
The $SU(3)$ breaking obtained here is slightly larger than the na\"{\i}ve 
expectation
as, in addition to the strange-quark mass effect, the $\la \bar ss\ra$ 
condensate also plays
an important role in the splitting.
\subsection{Hadronic 
and $\gamma\gamma$ widths}
\nin
$\b$ The hadronic and electromagnetic couplings of the lowest ground
states $S_2$ and $S_3$ have been estimated
using vertex sum rules.
The $S_2$ coupling to pair of pions in the chiral limit is \cite{BRAMON2}:
\beq
g_{S_2\pi^+\pi^-}\simeq \frac{16\pi^3}{3\sqrt{3}}\la 
\bar uu\ra\tau e^{M^2_2\frac{\tau}{2}}\simeq 2.46~ \mbox{GeV},
\eeq
for the typical value of $\tau\simeq 1$ GeV$^{-2}$, in good agreement with the
$SU(3)$ expectations. We thus deduce the width in Table 2.
 Using $SU(3)$ symmetry, one can also 
expect:
\beq
g_{S_2 K^+K^-}\simeq  \frac{1}{2}g_{S_2\pi^+\pi^-}.
\eeq
Analogous analysis for the $\gamma\gamma$ width leads to the predictions in Table 2.\\
$\b$ The estimates of the $\gamma\gamma$ and hadronic
widths of the radial excitations $S'_2$ and $S'_3$ are more uncertain. In so doing, we
use the phenomenological observations that the coupling of the radial excitation
increases as the ratio of the decay constants $r\equiv
f_{S_2}/f_{S'_2}$. Therefore, we expect:
\bea
\frac{\Gamma(S'_2\rar\gamma\gamma~ (\pi\pi))}
{\Gamma(S_2\rar\gamma\gamma~(\pi\pi))}
\approx r^2\ga\frac{M_{S'_2}}{M_{S_2}}\dr^{3(1)}
,
\eea
which, by taking $r\approx (M_{S'_2}/M_{S_2})^{(n=2\pm 1)}$, like in the pion 
and $\rho$ meson cases
\cite{SNB} gives the result in Table 2.\\
$\b$ To a first approximation, we expect that the decay of the $S'_2$ into
$4\pi$ comes mainly from the pair of $\rho$ mesons, while the one from
$\sigma_B\sigma_B$ (gluonia) is relatively suppressed like $\alpha_s^2$ 
using perturbative QCD arguments.
\section{Gluonium-quarkonium mass mixings}
\nin
This quantity can be obtained from the QSSR analysis of the off-diagonal
quark-gluon two-point correlator. It has been obtained for different
channels \cite{PAK,BRAMON,BRAMON2}. The results show that the mixing angle is tiny 
(less than 12$^\circ$) and justify a posteriori that the masses of the
observed gluonia are approximately given by the theoretical estimate of
the gluonia masses obtained without taking into account a such
term. Note that the mass-mixing between the 3- and 2-gluon bound states is also
small \cite{LATORRE}. 
\section{``Mixing-ology'' for the decay widths of scalar mesons}

\nin
In the following, we shall be concerned
with the mixing angle for the couplings, which, in
the same approach, is controlled by the off-diagonal
non-perturbative
three-point function which can (a priori) give a
large mixing angle. However, a QCD evaluation of this quantity is
quite cumbersome, such that in the following, we shall only fix the
decay mixing angle from a fit of the data.
\subsection{Mixing below 1 GeV and the nature of the $\sigma$ and $f_0(0.98)$}

\nin
We consider that the physically observed $f_0$ and $\sigma$ states
result from the two-component mixing of the $\sigma_B$ and $S_2\equiv
\frac{1}{\sqrt{2}}(\bar uu +\bar dd)$ unmixed bare states:
\bea
|f_0\ra &\equiv& -\sin\theta_S|\sigma_B\ra+\cos\theta_S|S_2\ra\nnb\\
|\sigma\ra &\equiv&~~\cos\theta_S|\sigma_B\ra+\sin\theta_S|S_2\ra .
\eea
Using the prediction:
$\Gamma({\sigma_B}\rar\gamma\gamma)\simeq (0.2\sim 0.3)~\mbox{keV},$
and the experimental width $\Gamma(f_0\rar\gamma\gamma)\approx 0.3$ keV,
one obtains \cite{BRAMON}:
\beq
\theta_S\approx (40\sim 45)^\circ,
\eeq
which indicates that, in this scheme,
the $f_0$ and $\sigma$ have a large amount of gluons in
their wave functions. This situation is quite similar to the case of 
the $\eta'$
in the pseudoscalar channel (mass given by its gluon component, but 
strong coupling to quarkonia).  
Using the previous value of $\theta_S$, the predicted value of $g_{S_2K^+K^-}$,
the approximate relation $g_{S_2K^+K^-}\simeq \frac{1}{2}g_{S_2\pi^+\pi^-}$,
and the almost universal coupling of the $\sigma_B$ to pairs of Goldstone 
bosons, one can deduce (in units of GeV): 
\bea\label{coupe}
 g_{f_0\pi^+\pi^-}&\simeq& (0.1\sim 2.6),\nnb\\ g_{f_0K^+K^-}&\simeq& 
-(1.3\sim 4.1)\nnb\\
 g_{\sigma\pi^+\pi^-}&\simeq& ~g_{\sigma K^+K^-}\simeq
(4\sim 5),
\eea
which can provide a simple explanation of the exceptional property of the $f_0$ 
(strong coupling to $\bar KK$ as observed in $\pi\pi$
and $\bar KK$ data \cite{PDG}), 
without appealing to the more exotic four-quark and $\bar KK$
molecules natures of these states
\footnote{A QSSR analysis of the $a_0(0.98)$ within a four-quark
scheme leads to too low a value of its $\gamma\gamma$ width as
compared with the data \cite{FOUR}.}.  
Using the previous predictions for the couplings,
and for $\theta_S$, we obtain the results in Table 3.
\subsection{Nature of the $f_0(1.37)$}
\nin
Among the observed widths of the $f_0(1.37)$, we shall mainly be concerned with the 
ones into $\gamma\gamma$ and $(4\pi)_S$ \cite{PDG,LANDUA}
showing that the $f_0(1.37)$ has amusingly the combined properties of the scalar
quarkonium $S'_2$ from its $\gamma\gamma$ width and of scalar gluonium $\sigma'_B$
from its decay into $(4\pi)_S$ through the pair of $\sigma$ states.
\nin
\begin{table*}[hbt]
\setlength{\tabcolsep}{0.7pc}
\caption{ Predicted decays of the observed scalar mesons}
\begin{tabular}{ c c c c c c c}
\hline 
Name&$\pi^+\pi^-$[MeV]&$K^+K^-$[MeV]&$\eta\eta$
[MeV]&$\eta\eta'$[MeV]&$(4\pi^0)_S$[MeV]&$\gamma\gamma$[keV]\\
&&&&&&\\
\hline 
$f_0(0.98)$&$ 0.2\sim 134$&Eq. (\ref{coupe})&$$&&&$\approx 0.3$\\
&&&&&&(exp)\\
$\sigma(0.75\sim 1)$&$300\sim 700$&Eq. (\ref{coupe})&$SU(3)$&&$$&$0.2\sim 0.5$\\
&&&&&&\\
$f_0(1.37)$&$22\sim 48$&$\approx 0$&$\leq 1.$&$\leq 2.5$&$150$&$\leq 2.2$\\
&&(exp)&&&&\\
$f_0(1.5)$&25&$3\sim 12$&$1\sim 2$&$\leq 1.$&$68\sim 105$&$\leq 1.6$\\
&(exp)&&&&(exp)&\\
$f_J(1.71)$&$\approx 0$&$112\pm 50$&$SU(3)$&&$\approx 0$&$1.1\pm 0.5$\\
\hline
\end{tabular}
\end{table*}
\nin
\subsection{3x3 mixing and nature of the $f_0(1.5)$}
\nin
In order to explain the nature of the $f_0(1.5)$, we need to consider the
3x3 mixing matrix in Eq. (\ref{matrix}).
\begin{table*}[hbt]
\catcode`?=\active \def?{\kern\digitwidth}
\beq\label{matrix}
\left(
\begin{array}{c}
f_0(1.37)\\
f_0(1.50)\\
f_0(1.60)\\
\end{array}
\right)
\approx
\left(
\begin{array}{ccc}
0.01\sim 0.22&-(0.44\sim 0.7)&0.89\sim 0.67\\
0.11\sim 0.16&0.89\sim 0.71&0.43\sim 0.69\\
-(0.99\sim 0.96)&-(0.47\sim 0.52)&0.14\sim 0.27\\
\end{array}
\right)
\left(
\begin{array}{c}
\sigma'_B(1.37)\\
S_3(1.47)\\
G(1.5)
\end{array}
\right)
\eeq
\end{table*}
\nin
The mixing angles in the first line of the matrix have been fixed by using the 
negligible $\bar 
KK$ and $\pi\pi$
widths of the $f_0(1.37)$ given in Table 3. For the second line of the matrix, 
we use the observed 
width of the $f_0(1.5)$ into $\pi\pi$. 
The first (resp. second) numbers in the matrix correspond to the case of large 
(resp. small) widths from the data. From the previous schemes, 
we deduce the predictions in Table 3
{\footnote{The present data favour negative values of the $f_0\eta\eta$,
$f_0\eta'\eta$ and $f_0KK$ couplings.}.
The orthogonal $f_0(1.6)$ state is too broad for being
considered as a resonance
 \footnote{In our scheme, the state observed either by GAMS or by
Crystal Barell should be compared with the narrow $f_0(1.5)$.}.
Despite the crude approximation used and the inaccuracy of the
predictions, these results are
in good agreement with the data (especially from the
Crystal Barrel collaboration), and suggest that the 
observed $f_0(1.37)$ and $f_0(1.5)$ come from a maximal mixing
between the gluonia ($\sigma'_B$ and $G$) and the quarkonium $S_3$
states. The mixing of the $S_3$ and $G$
with the quarkonium $S'_2$, which we have neglected compared with
the $\sigma'_B$, can restore the small discrepancy with the data.
One should notice, as already mentioned, that the state seen 
by GAMS is more likely the unmixed
gluonium state $G$ (dominance of the
$4\pi$ and $\eta\eta'$ decays,
as emphasized earlier in NV), which can be due to some specific
features of the production at the GAMS experiment, but not present
in the Crystal Barrel and Obelix ones.
\subsubsection{Nature of the $f_J(1.71)$}
\nin
The
narrow $f_J (1.7)$ observed to
decay into $\bar KK$ with a width of the order $(100\sim 180)$ MeV
can be essentially composed by the radial excitation 
$S'_3(1.7\sim 2.4)$ GeV of the $S_3(\bar ss)$, as they have
about the same width into $\bar KK$ (see Table 2). This feature can
also explain
the smallness of the $f_J(1.7)$ width into $\pi\pi$ and $4\pi$.
Our predictions of the $f_J(1.71)$ width can agree with
the result of the Obelix collaboration \cite{LANDUA}, while
its small decay width into $4\pi$ is in agreement with the
best fit of the Crystal Barrel collaboration 
\cite{LANDUA}, which is consistent with
the fact that the $f_0(1.37)$ likes to decay into 4$\pi$.
However, the broad $f_0(1.6)$ and the $f_J(1.71)$ can presumably
interfere destructively for giving the dip around 
$1.5\sim 1.6$ GeV seen in 
the $\bar KK$ mass distribution
from the Crystal Barrel and $\bar pp$
annihilations at rest. 
\subsubsection{Comparison with other scenarios}

\nin
Though 
the relative amount of glue for the $f_0(1.37)$ and $f_0(1.5)$
is about the same here and in \cite{CLOSE2} , one should notice
that, in our case, the $\pi\pi$ partial width of these mesons come mainly
from the $\sigma'_B$, a glue state coupled strongly to the
quark degrees of freedom, like the $\eta'$ of the $U(1)_A$ anomaly,
while in \cite{CLOSE2}, the $S_2$ which has a mass higher than
the one obtained here plays an essential role in the mixing.
Moreover, the $f_J(1.71)$ differs significantly in the two approaches, as
here, the $f_J(1.71)$ is mainly the $\bar ss$ state $S'_3$, 
while in \cite{CLOSE2},
it has a significant gluon component. In the present approach, the eventual
presence of a large gluon component into the $f_J(1.71)$ wave function
can only come from the mixing
with the broad $f_0(1.6)$ and with
the radial excitation of the gluonium $G$(1.5), which mass
is expected to be around 2 GeV as suggested by the QSSR
analysis. However, the apparent absence
of the $f_J(1.71)$ decay into 4$\pi$ from Crystal Barrel data may not
favour such a scenario.
\section{Conclusions}
\nin
 We have reviewed:\\ 
$\b$ The QCD spectral sum rule (QSSR) predictions
of the masses and
decay constants of gluonia, and given some interpretations of the
nature of the observed $\zeta(2.2)$ and $E/\iota(1.44)$ mesons (Table 1).\\
\nin
$\b$ Some low energy theorems (LET) and vertex sum rule estimates
of the widths of the scalar gluonia quarkonia (Table 2).\\
\nin
$\b$ Some maximal
quarkonium-gluonium mixing schemes, for explaining the
complex structure and decays of the observed scalar mesons (Table 3).\\
The good
agreements between the theoretical predictions and the data are 
encouraging.

\section*{DISCUSSIONS}
\nin
{\bf M. Loewe}, Santiago (Chili)\\
{\it What are the values of the condensates you used in your estimation
of the masses of gluonium states?}
\\
\nin
{\bf S. Narison}\\
{\it I use the new values of the gluon condensates (see text)
obtained from the $e^+e^-\rar I=1$ hadrons data ($\tau$-like sum rule) and
the heavy quark-mass splittings.}\\
{\bf G. Schuler}, CERN (Geneva)\\
{\it You predict a 3-gluon state at 3.1 GeV:\\
i) Do you have any ideas about its decay and where one can look for it?\\
ii) Can this state any of the $J/\psi$ decays ?}\\
\nin
{\bf S. Narison}\\
{\it i) This scalar state can preferably decay into the $U(1)$ channels $\eta'\eta',~\eta'\eta$
and $\eta\eta$. It can also decay into $4\pi$ in a S-wave through the pair 
of broad scalar state $\sigma$. Its hadronic coupling can be suppressed by a factor
$\alpha_s$ relatively to the analogue scalar gluonium $f_0(1.5)$ formed by two gluons. If,
it also couples weakly to $\pi\pi$ and $KK$, its 
experimental detection will be difficult. \\
ii) I do not think so, as, in addition,
 its production is strongly suppressed by phase space factors.

\begin{thebibliography}{999}
\bibitem{SN}S. Narison, {\it Masses, decays and mixings of gluonia in QCD},
hep-ph/9612457(1996) ({\it Nucl. Phys.} {\bf B} (in press)).
\bibitem{FRITZ}M. Gell-Mann, {\it Acta Phys. Aust. Suppl} {\bf 9} 
(1972) 733;
H. Fritzsch and M. Gell-Mann, {\it XVI ICHEP Conf.}, Chicago, {\bf Vol 2} (1972) 135; 
H. Fritzsch and P. Minkowski, 
{\it Nuovo Cimento}
{\bf 30A} (1975) 393.
\bibitem{SVZ}M.A. Shifman, A.I. Vainshtein and V.I. Zakharov,
{\it Nucl. Phys.} {\bf B147} (1979) 385, 448.
\bibitem{SNB}S. Narison, {\it QCD spectral sum rules}, Lecture Notes in
Physics, {\bf Vol. 26} (WSC 1989);
{\it Recent progress in QSSR} (in preparation).
\bibitem{VENEZIA}S. Narison and G. Veneziano, {\it Int. J. Mod. Phys} 
{\bf A4, 11} (1989) 2751.
\bibitem{SNL} S. Narison, {\it Phys. Lett.} {\bf B361} (1995) 121.
\bibitem{SNH} S. Narison, {\it Phys. Lett.} {\bf B387} (1996) 162.
\bibitem{ZAK} R. Akhoury and V. I. Zakharov, 
{\it Nucl. Phys. (Proc. Suppl.)} {\bf B, A54} (1997);
{\it QCD97}, Montpellier (1997) 
and references therein.
\bibitem{SETTLES}E. Braaten, S. Narison and A.
Pich, {\it Nucl. Phys.} {\bf B373} (1992) 581; 
R. Settles, {\it QCD97},
Montpellier (1997) and references therein.
\bibitem{NSVZ}V.A. Novikov et al., {\it Nucl. Phys.} {\bf B191} (1981) 301.
\bibitem{BETHKE}S. Bethke, 
{\it Nucl. Phys. (Proc. Suppl.)} {\bf B, A54} (1997);
M. Schmelling,
 {\it IHEP}, Varsaw
(1996).
\bibitem{BB}
J.S. Bell and R.A. Bertlmann, {\it Nucl. Phys.} {\bf B177} (1981) 218;
{\bf B227} (1983) 435.
\bibitem{SR}S. Narison and E. de Rafael, {\it Phys. Lett.} 
{\bf B103} (1981) 87.
\bibitem{LAUNER} R.A. Bertlmann, G. Launer and E. 
de Rafael,
{\it Nucl. Phys.} {\bf B250} (1985) 61.
\bibitem{TEPER}M. Teper, {\it IHEP}, Jerusalem (1997)
and references therein.
\bibitem{WEST} G. West, {\it Nucl. Phys. (Proc. Suppl.)} {\bf B, A54} (1997).
\bibitem{BRAMON} E. Bagan, A. Bramon and S. Narison,
{\it Phys. Lett.} {\bf B196} (1987) 203. 
\bibitem{LANDUA}For reviews, see e.g.: M. Faessler, {\it IHEP}
Jerusalem (1997); R. Landua, {\it IHEP}, 
Varsaw (1996); J.P. Stroot, {\it QCD97},
Montpellier (1997); ibid. N. Djaoshvili and L. Montanet; 
U. Gastaldi, Legnaro preprint LNL-INFN (Rep) (1997).
\bibitem{NSVZ2} V.A. Novikov et al. {\it Nucl. Phys.} {\bf B165} (1980) 67.
\bibitem{WITTEN} E. Witten, {\it Nucl. Phys.} {\bf B156} (1979) 269;\\
G. Veneziano, {\it Nucl. Phys.} {\bf B159} (1979) 213.
\bibitem{PDG}PDG: R.M. Barnett et al., {\it Phys. Rev.} {\bf D54} (1996) 1.
\bibitem{GILMAN}F. Gilman and R. Kauffman, {\it Phys. Rev.} {\bf D36} (1987) 
2761;
L. Montanet, {\it Non-Perturbative Methods Conf.}, Montpellier, (WSC 1985). 
\bibitem{GERS}S.S. Gershtein, A.A. Likhoded and Y.D. Prokoshkin,
{\it Z. Phys.} {\bf C24} (1984) 305.
\bibitem{ELLIS}R.J. Crewther, {\it Phys. Rev. Lett.} {\bf 28} (1972) 1421;
J. Ellis and M.S. Chanowitz, {\it Phys. Lett.} 
{\bf B40} (1972) 397; {\it Phys. Rev.} {\bf D7} (1973) 2490.
\bibitem{CHAN}M.S. Chanowitz, {\it Proc. of the VI Int. Workshop on 
$\gamma$-$\gamma$ collisions}, (WSC 1984).
\bibitem{SNM} S. Narison, {\it Phys. Lett.} {\bf B358} (1995) 113.
\bibitem{BRAMON2}A. Bramon and S. Narison, {\it Mod. Phys. Lett.} 
{\bf A4} (1989) 1113.
\bibitem{PAK}S. Narison, N. Pak and N. Paver, {\it Phys. Lett.} {\bf B147}
(1984) 162.
\bibitem{LATORRE}J.I. Latorre, S. Paban and S. Narison,
{\it Phys. Lett.} {\bf B191} (1987) 437.
\bibitem{FOUR}S. Narison, {\it Phys. Lett.} {\bf B175} (1986) 88.
\bibitem{CLOSE2}F.E. Close, {\it Proc. LEAP96}, Dinkelsbul (1996) 
(hep-ph/9610426) and references therein. 
%
\end{thebibliography}
\end{document}